\documentclass[nature]{wlscirep}
\usepackage{graphicx}
\usepackage{dcolumn}
\usepackage{bm}
\usepackage{threeparttable}
\usepackage{rotfloat}
\usepackage{float}
\usepackage{color}
\usepackage{amsmath}
\def\degree{${}^{\circ}$}

\begin{document}

\title{Nonsymmorphic symmetry protected node-line semimetal in the trigonal YH$_{3}$}
\author[1]{Dexi Shao}
\author[1]{Tong Chen}
\author[1]{Qinyan Gu}
\author[1]{Zhaopeng Guo}
\author[1]{Pengchao Lu}
\author[1*]{Jian Sun}
\author[1]{Li Sheng}
\author[1]{Dingyu Xing}

\affil[1]{National Laboratory of Solid State Microstructures,
School of Physics and
Collaborative Innovation Center of Advanced Microstructures,
Nanjing University, Nanjing 210093, China}
\affil[*]{jiansun@nju.edu.cn}

\begin{abstract}
Using ab initio calculations based on density-functional theory and effective model analysis,
we propose that the trigonal YH$_{3}$ (Space Group: P$\bar{3}$c1) at ambient pressure
is a node-line semimetal when spin-orbit coupling (SOC) is ignored.
This trigonal YH$_{3}$ has very clean electronic structure near Fermi level
and its nodal lines locate very closely to the Fermi energy,
which makes it a perfect system for model analysis.
Symmetry analysis shows that the nodal ring in this compound is protected by the glide-plane symmetry,
where the band inversion of $|Y^{+},d_{xz}>$ and $|H1^{-},s>$ orbits at $\Gamma$ point
is responsible for the formation of the nodal lines.
When SOC is included, the line nodes are prohibited by the glide-plane symmetry,
and a small gap ($\approx$ 5 meV) appears,
which leads YH$_{3}$ to be a strong topological insulator with Z$_{2}$ indices (1,000).
Thus the glide-plane symmetry plays an opposite role in the formation of the nodal lines in cases without and with SOC.
As the SOC-induced gap is so small that can be neglected,
this P$\bar{3}$c1 YH$_{3}$ may be a good candidate for experimental explorations
on the fundamental physics of topological node-line semimetals.
We find the surface states of this P$\bar{3}$c1 phase are somehow unique
and may be helpful to identify the real ground state of YH$_{3}$ in the experiment.
\end{abstract}

\maketitle

\section*{Introduction}
Topological semimetals (TSMs) have attracted great attention for
both theoretical interests and experimental applications in recent years.
Different from time-reversal symmetry (TRS) protected
Z$_{2}$ topological insulators (TIs)~\cite{TIs-Rmp-2010-Hasan,TIs-with-inversion-2011-(1057)FuLiang}
which are insulating in the bulk,
TSMs are materials where the conduction and the valence bands
cross with each other at certain locations in the Brillouin Zone (BZ).
Usually, the band crossing is protected by certain symmetries, i.e.,
perturbations on the Hamiltonian which respect the symmetries
can not break the crossing.
Recently, several types of TSMs have been proposed
to investigate the fermion-like excitations,
including Dirac fermions~\cite{DS-AB3-Wang-2012,3DDS-Cd3As2-Wang-2013,3D-BiO2,3dssm-design},
Weyl fermions~\cite{Wan2011-weyl,HgCrSe-2011,TaAs-exper,Xu2015-TaAs}
and nodal lines~\cite{graphene-network,Cu3N-newZ2,Cu3PdN,Bian2016-PbTaSe2,Bzdu2016Nodal,nodalline-alkali-2016}.
These compounds are named as: Dirac semimetals (DSMs), Weyl semimetals(WSMs) and node-line semimetals (NLSMs), respectively.

Up to now, there have been a lot of reports about the progress in DSMs and WSMs,
for example, 
3D Dirac semimetals have recently been identified experimentally in Cd$_{3}$As$_{2}$~
\cite{Cd3As2-Liu-NM-2014,Observation-Cd3As2-NC-XuSuYang-2014,
Exp-realization-DS-Cava-2014,Cd3As2:Evi-surface-state-Yi-2013,
Landau-quan-CdAs-NM-Jeon-2014}
and Na$_{3}$Bi systems~\cite{Exp-verify-Na3Bi-Sci-2014,Exp-fermi-arc-Sci-XSY-2015}.
Similarly, TaAs~\cite{TaAs-exper,TaAs-weyl-obser,TaAs-weylnode-obs,spintexture-TaAs},
NbAs~\cite{NbAs-weyldiscovery}, TaP~\cite{TaP-2016observation},
WTe$_2$~\cite{WTe2-fermiarc-2016,WTe2-fermiarc2-2016}, and MoTe$_2$~\cite{MoTe2-2016experim}
etc are verified to be WSMs experimentally in recent years.
Different from DSMs and WSMs, in which the conduction and valence bands touch at discrete points,
the crossings of NLSMs form a closed loop in the BZ.
Although many candidates of NLSMs have been proposed
and much efforts has been made to investigate them,
the corresponding progress in the experiment is slow,
because an open surface usually breaks the inversion
or some mirror symmetries which are important to the formation of nodal lines.~\cite{nlsm-fc-rapidcom}

Materials experimentally confirmed (or partially confirmed) to host nodal line
include Be metal~\cite{nodalline-alkali-2016},
ZrSiS~\cite{ZrSiS-exp-2016}, PbTaSe$_2$~\cite{Bian2016-PbTaSe2} and ZrSiSe/ZrSiTe~\cite{ZrSiSeTe-exp-2016}.
Therefore, theoretical predictions on more candidates of node-line semimetal are still in demand.
It is well known that symmetries play important roles
in identifying various of TIs and topological superconductors (TSCs)~
\cite{class-TI-Tsc-2008,TCI-FuLiang-2011,SG-class-2013,TCI-SG-2013,TCI-TSC-2014}.
In fact, symmetries are also important in classifying NLSMs, for instance,
three types of NLSMs protected by different symmetries have been proposed:
(a) mirror symmetry protected NLSMs~\cite{HgCrSe-2011,classify-msym-2014-chiu,Bian2016-PbTaSe2,nodalline-alkali-2016},
(b) coexistence of TRS and inversion symmetry (IS) protected NLSMs~\cite{graphene-network,Cu3N-newZ2,Cu3PdN,nlsm-fc-rapidcom} and
(c) nonsymmorphic symmetry protected NLSMs\cite{Bzdu2016Nodal,nlsm-fc-rapidcom}.

Hydrides is a large class of materials and has been extensively investigated in many aspects,
including energy storage~\cite{H-storage} and superconductivity~\cite{BaH-2007,SnH4-2007,SiH3-2010,IH-2015,HTSC-cui2017}, etc.
Since Ashcroft proposed that high $T_{c}$ superconductivity
can be obtained in hydrogen and hydrides under high pressure,~\cite{metalH,metalH-alloy}
many hydrides have been investidated.
Yttrium-hydrogen system becomes interesting due to the same reason. For instance,
a fcc YH$_{3}$ has been predicted to be a superconductor with $T_{c} \sim$ 40 K at 17.7 GPa.~\cite{YH3-sc}
Later work predicts that two new yttrium hydrides, i.e., YH$_4$  and YH$_6$, are also superconductors
with $T_{c} \sim$ 84 - 95 K and $T_{c} \sim$ 251 - 264 K at 120 GPa, respectively.~\cite{YH4-6-SC}
Very recently, YH$_{10}$ in the space group of both $Im\bar{3}m$ and $Fm\bar{3}m$ has been predicted
to be a room-temperature superconductor under very high pressure.~\cite{YH10-liu,YH10-ma}

Though many works about superconductivity of yttrium-hydrogen system under pressure have been implicated,
very few explorations on the topological electronic properties of hydrides have been reported so far~\cite{BiX-monolayer,soc-mat,2d-NL}.
In this work, we predicted that YH$_{3}$ in the space group of P$\bar{3}$c1 at ambient pressure
is a node-line semimetal when spin-orbit coupling (SOC) is ignored.
Especially, the YH$_{3}$ system we studied has extremely clean electronic structures near the Fermi level;
i.e., there are no other pockets.
The energy of the crossing points along the nodal loop varies within a very small energy range, from around -5 to 35 meV.
Therefore, this nodal loop is very "flat" in the energy and momentum space,
which makes YH$_{3}$ a perfect model system for NLSMs.
In general, NLSMs without SOC will transform into either insulators, DSMs, WSMs or even double NLSMs when SOC is considered,
which is much related to the symmetries in the corresponding systems.~\cite{NLSM-sum}
While in this work, when SOC is included, the three nodal lines around $\Gamma$ point will be gapped out with a small gap ($\approx$ 5 meV),
making YH$_{3}$ a topological insulator with Z$_{2}$=(1,000).
Nevertheless, further calculation shows that the gap induced by SOC along the nodal ring is very small (about 5 meV),
which indicates that the effect of SOC is negligible and the characteristic of the nodal ring can be preserved.

\section*{Methods}
Calculations of the band structures are performed using the full-potential linearized
augmented plane-wave (FP-LAPW) method~\cite{LAPW,LAPW-prb} implemented in the WIEN2k~\cite{wien2k} package.
We use $13\times13\times11$ k-mesh for the BZ sampling
and -7 for the plane wave cut-off parameter R$_{MT}$K$_{max}$ for the electronic structure calculation,
where the R$_{MT}$ is the minimum muffin-tin radius
and K$_{max}$ is the plane-wave vector cut-off parameter.
SOC is taken into consideration by a second-variation method~\cite{soc}.
The tight-binding models are constructed with
the maximally localized Wannier functions (MLWFs) method~\cite{mlwfs-1997,MLWFs-RMPhys,Mostofi2007wannier90},
the corresponding hopping parameters are determined from
the projections of the bulk Bloch wave functions.
The projected surface states are calculated using surface Green's function in the semi-infinite system~\cite{Qui-iter-sche,SurfaceGF}.

\section*{Results and discussions}
{\bf The crystal structure of YH$_{3}$.}
Historically, three different phases of YH$_{3}$ have been reported to be the ground state potentially.
Two of them are experimentally favoured with trigonal P$\overline{3}$c1 and hexagonal P6$_{3}$cm symmetry~\cite{YD3-npd,YD3-prb,Pereierls-YH3-prl,LaH,P-3c1-agree,P-3c1-2006},
while the third candidate is in the space group of P6$_{3}$ which was predicted theoretically.~\cite{YH3-P63-prl}
It seems that the ground state of YH$_{3}$ at ambient pressure
is still under debate because the three candidates
have very tiny total energy difference (~0.001 eV/f.u.) from theoretical point of view.
First, the hexagonal P6$_{3}$ structure is only theoretically proposed
and seems to disagree with the neutron-diffraction results~\cite{YH3-P63-prl-neu1,YH3-P63-prl-neu2}.
Second, later neutron-diffraction experiments~\cite{P-3c1-2006,YH3-P-3c1-nat} identify
that the P$\bar{3}$c1 structure is stable from the ambient pressure up to 12GPa.
Thus, in the following, we only focus on the phase of YH$_{3}$ with the P$\bar{3}$c1 symmetry.

\begin{figure}[th]
\begin{center}
\includegraphics[width=0.75\textwidth]{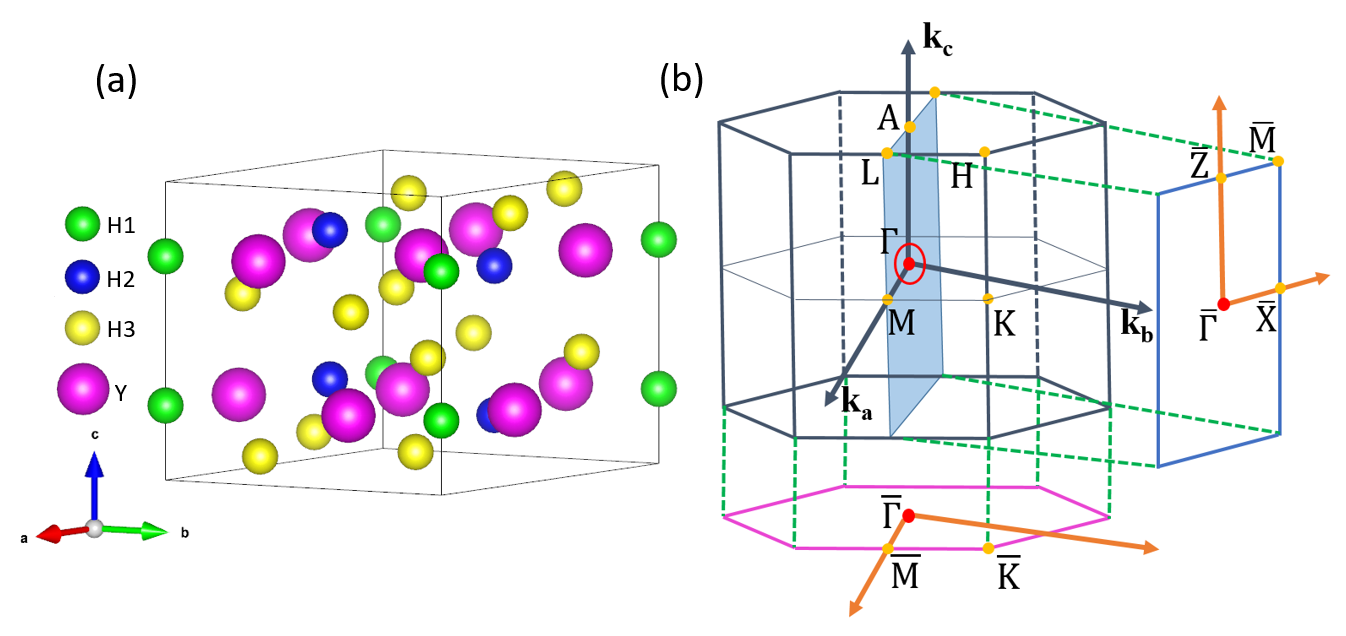}
\caption{%
  (a)Crystal structure of YH$_{3}$ at the ambient pressure with P$\bar{3}$c1 symmetry.
H1, H2 and H3 atoms occupy the 2a (0,0,$\frac{1}{4}$), 4d ($\frac{1}{3}$,$\frac{2}{3}$,0.181)
and 12g (0.348,0.025,0.093) sites, respectively,
while Y atoms lie at the 6f (0.336,0,$\frac{1}{4}$) sites.
(b) The corresponding BZ and its projection onto the (010) direction.
The red ring on the shadow plane surrounding the $\Gamma$ point represents the node-line structure in the BZ.
}
\label{fig:YH3-P-3c1}
\end{center}
\end{figure}

The crystal structure and corresponding BZ of P$\bar{3}$c1 (Space Group No. 165) YH$_{3}$ is shown in Fig.~\ref{fig:YH3-P-3c1}(a)
and Fig.~\ref{fig:YH3-P-3c1}(b) respectively.
We use the experimental lattice parameters from literature~\cite{P-3c1-2006}
in our calculations, which
are listed in Table.~\ref{table:str-parameters}.\par

\begin{table}
\centering
\caption{The experimental lattice parameters of YH$_{3}$ with P$\bar{3}$c1 symmetry.\cite{P-3c1-2006}}
   \begin{tabular}{l c c c c r}%
    \hline\hline
          \textbf{phase}             & \textbf{a=b}               & \textbf{c}         &  \textbf{$\alpha=\beta$} &  \textbf{$\gamma$}\\
           P$\overline{3}$c1         & 6.359 (${\AA}$)            & 6.607 (${\AA}$)    &     90\degree            &   120\degree\\
     \hline\hline
  \end{tabular}
\label{table:str-parameters}
\centering
\end{table}

\begin{figure}[th]
\begin{center}
\includegraphics[width=0.75\textwidth]{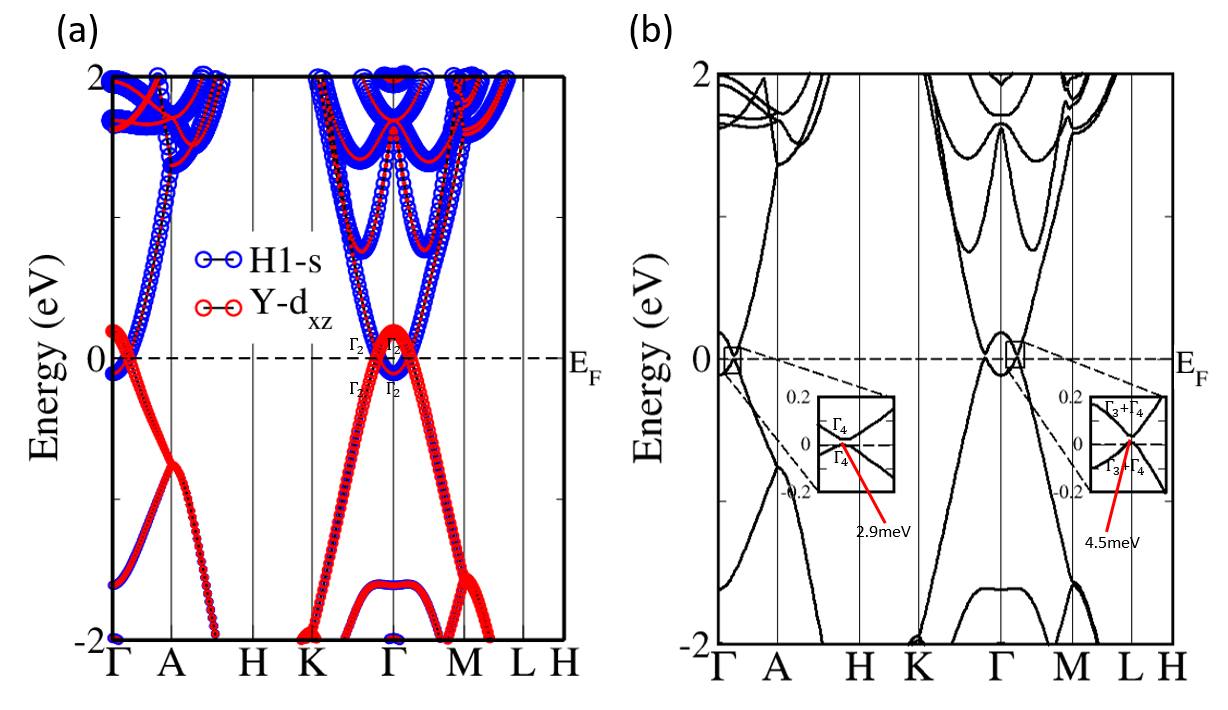}
\caption{%
  (a) Corresponding fat-band structure of YH$_{3}$ in the space group of P$\bar{3}$c1 without SOC.
  The bands between CBM and VBM is gapped along $K\rightarrow \Gamma$ as the irreducible representations have showed.
  (b) Corresponding band structure of YH$_{3}$ in the space group of P$\bar{3}$c1 with SOC.
}
\label{fig:YH3-P-3c1-icsd.bands}
\end{center}
\end{figure}

{\bf Band structures without SOC and the corresponding model analysis.}
From the band structures of P$\bar{3}$c1 YH$_{3}$ without SOC shown in Fig.~\ref{fig:YH3-P-3c1-icsd.bands}(a),
we can find three Dirac crossings composed of the conduction band minimum (CBM) and the valence band maximum (VBM)
near $\Gamma$ along the high-symmetry path in the BZ at the first sight.
Detailed first-principle calculations indicates that Dirac crossings
lying in the plane $m_{{\Gamma}-M-L-A}$ is protected by the $G_{x}$ symmetry,
while Dirac crossing along $K \rightarrow \Gamma$ is not symmetry-protected
(both the irreducible representations of CBM and VBM are $\Gamma_{2}$)
i.e., hybridization between the CBM and the VBM will open a gap in-between.
Further calculations indicates that the CBM and the VBM
are contributed mainly by Y-d$_{xz}$ and H1-s orbits, respectively,
as the fat-band showed in Fig.~\ref{fig:YH3-P-3c1-icsd.bands}(a).
The band inversion (not caused by SOC) of Y-d$_{xz}$ and H1-s at $\Gamma$ point leads to a gap of 0.302 eV.

To give more insights of the nodal lines surrounding the $\Gamma$ point,
we have established an effective Hamiltonian
by $\mathbf{k}\cdot\mathbf{p}$ method.
Taking the crystal symmetry and TRS into consideration,
the effective Hamiltonian can be written as follows:
\begin{equation}\label{eq:eps}
\begin{split}
&H(\vec{k})=g_{0}(\vec{k})\tau_{0}+g_{x}(\vec{k})\tau_{x}+g_{z}(\vec{k})\tau_{z}\\
&g_{0}(\vec{k})=M_{0}-B_{0}(k_{x}^{2}+k_{y}^{2})-C_{0}k_{z}^{2}\\
&g_{x}(\vec{k})=A(k_{x}^{3}-3k_{x}k_{y}^{2})\\
&g_{z}(\vec{k})=M_{z}-B_{z}(k_{x}^{2}+k_{y}^{2})-C_{z}k_{z}^{2}.
\end{split}
\end{equation}
Here, the $\tau_{x}$ and $\tau_{z}$ are Pauli matrices, $\tau_{0}$ is a $2\times 2$ unit matrix.
This system has both TRS and IS, thus, the component of $\tau_{y}$ must be zero~\cite{graphene-network}.
We can obtain the eigenvalues of the two-level system by diagonalizing the $2\times 2$ effective Hamiltonian
and the results are $E(\vec{k})=g_{0}(\vec{k})\pm \sqrt{g_{x}^{2}+g_{z}^{2}}$.
Band crossings of the nodal lines will occur when $g_{x}=g_{z}=0$.
It's clear that $g_{z}(\vec{k})=0$ gives us $M_{z}B_{z}>0$ and $M_{z}C_{z}>0$.
We find that $M_{z}B_{z}>0$ and $M_{z}C_{z}>0$ are exactly the condition of band inversion.
Furthermore, $g_{x}(\vec{k})=0$ confines the band crossings
of the node-line in the $k_{x}=0,\pm \sqrt{3}k_{y}$ planes,
i.e., there are three nodal rings surrounding the $\Gamma$ point and lying in mirror-invariant planes $m_{{\Gamma}-M-L-A}$,
as shown with the shadow sector in Fig.~\ref{fig:YH3-P-3c1}(b).
It's obvious that these three nodal rings are related to each other by $R_{3z}$ symmetry.

{\bf Band structures with SOC and the corresponding model analysis.}
When SOC is considered, band crossings of the three nodal lines will disappear,
as the corresponding band structure with SOC shown in Fig.~\ref{fig:YH3-P-3c1-icsd.bands}(b).
We will further explain the above-mentioned phenomenon in the following.
Taking SOC into account,
spin and orbital angular momentum are coupled together,
which generates a group of new eigenstates with certain total angular momentum quantum numbers.
Then we can mark these new eigenstates of the CBM and VBM as
$|H1_{s}^{-},\pm\frac{1}{2}>$,
and $|Y_{d_{xz}}^{+},\pm\frac{1}{2}>$.
Here subscripts s and d$_{xz}$ denote corresponding orbits consisting of the new eigenstates and
the superscripts $\pm$ represent the parities of corresponding eigenstates, respectively.

According to the analysis of irreducible representations and projected orbits,
the CBM and VBM at the $\Gamma$ point
(denoted as $\Gamma_{4}^{-}$ and $\Gamma_{4}^{+}$) are mainly composed of
$|H1_{s}^{-},\pm\frac{1}{2}>$ and $|Y_{d_{xz}}^{+},\pm\frac{1}{2}>$ basis, respectively.
If we arrange the 4 basis in the order of
$|H1_{s}^{-},\frac{1}{2}>$,$|H1_{s}^{-},-\frac{1}{2}>$,
$|Y_{d_{xz}}^{+},\frac{1}{2}>$,$|Y_{d_{xz}}^{+},-\frac{1}{2}>$,
and then take the time-reversal and $D_{3d}$ point-group symmetries at the $\Gamma$ point into consideration,
we can give the character table of $\Gamma$ matrices and the polynomials of momentum $\vec{k}$
for this system as shown in Table.~\ref{table:kpolynomials-soc}.

\begin{table}
\centering
\caption{The character table for the P$\bar{3}$c1 YH$_{3}$.}
   \begin{tabular}{c c c c}%
       \hline\hline
          \textbf{$\Gamma$}               &      Representation                       &   T/P     & $\vec{k}$ \\
           $\Gamma_{1}$                   &    $\tilde{\Gamma}_{1}^{+}~/~A_{1g}$         &    +      & $1,k_{x}^{2}+k_{y}^{2},k_{z}^{2}$ \\
           $\Gamma_{2}$                   &    $\tilde{\Gamma}_{2}^{-}~/~A_{2u}$       &    -      & $k_{z},k_{z}^{3},3k_{x}^{2}k_{y}-k_{y}^{3}$ \\
           $\Gamma_{3}$                   &    $\tilde{\Gamma}_{1}^{-}~/~A_{1u}$       &    -      & $k_{x}^{3}-3k_{x}k_{y}^{2}$ \\
           $\{\Gamma_{4},\Gamma_{5}\}$    &    $\tilde{\Gamma}_{3}^{-}~/~E_{u}$        &    -      & $\{k_{x},k_{y}\},\{k_{x}^{3}+k_{x}k_{y}^{2},k_{x}^{2}k_{y}+k_{y}^{3}\}$ \\
     \hline\hline
  \end{tabular}
      \label{table:kpolynomials-soc}
\end{table}

From Table.~\ref{table:kpolynomials-soc}, our model Hamiltonian yields as
\begin{equation}\label{eq:eps}
\begin{split}
H(\vec{k})&=\epsilon_{0}(\vec{k})+\sum_{i=1}^{5}f_{i}(\vec{k})\Gamma_{i}\\
          &=\epsilon_{0}(\vec{k})+
\left(
  \begin{array}{cccc}   %
  M(\vec{k})         &               0               &                A(\vec{k})        &           C(\vec{k})                \\  %
      0              &          M(\vec{k})           &                C^{*}(\vec{k})    &           B(\vec{k})         \\  %
  A^{*}(\vec{k})     &          C(\vec{k})           &               -M(\vec{k})        &             0                \\  %
  C^{*}(\vec{k})     &          B^{*}(\vec{k})       &                  0               &          -M(\vec{k})         \\  %
  \end{array}
\right)
\end{split}
\end{equation}
which describes the dispersion of the CBM and VBM around the $\Gamma$ point.
Here we use the following $\Gamma$ matrices:
\begin{equation}\label{eq:eps}
\begin{split}
&\Gamma_{1}=\sigma_{3}\otimes\tau_{0}~~~~\Gamma_{2}=\sigma_{1}\otimes\tau_{3}~~~~\Gamma_{3}=\sigma_{2}\otimes\tau_{0}\\
&\Gamma_{4}=\sigma_{1}\otimes\tau_{1}~~~~\Gamma_{5}=\sigma_{1}\otimes\tau_{2},
\end{split}
\end{equation}
which satisfy the Clifford algebra $\{\Gamma_{a},\Gamma_{b}\}=2\delta_{ab}$.
While the other ten $\Gamma$ matrices are given by
$\Gamma_{ab}=\frac{1}{2i}[\Gamma_{a},\Gamma_{b}]$.
Presence of both TRS and IS will forbid the existence of
these ten $\Gamma_{ab}$ terms in our model Hamiltonian.
In Equation~(2), $\epsilon_{0}(\vec{k})=D_{0}-m_{0}k_{z}^{2}-n_{0}(k_{x}^{2}+k_{y}^{2}), M(\vec{k})=D_{1}-m_{1}k_{z}^{2}-n_{1}(k_{x}^{2}+k_{y}^{2}),A(\vec{k})=D_{2}k_{z}+E_{2}k_{z}^{3}+F_{2}(3k_{x}^{2}k_{y}-k_{y}^{3})-iD_{3}(k_{x}^{3}-3k_{x}k_{y}^{2})$, $B(\vec{k})=-D_{2}k_{z}-E_{2}k_{z}^{3}-F_{2}(3k_{x}^{2}k_{y}-k_{y}^{3})-iD_{3}(k_{x}^{3}-3k_{x}k_{y}^{2})$ and
$C(\vec{k})=D_{45}k_{-}+E_{45}(k_{x}^{2}+k_{y}^{2})k_{-}$ with $k_{-}=k_{x}-ik_{y}$.

From the model Hamiltonian given in Equation.~(2)
together with the band structure shown in Fig.~\ref{fig:YH3-P-3c1}(b),
we can draw some conclusions as the following.
First of all, $\epsilon_{0}(\vec{k})$ will break the particle-hole symmetry
for the CBM and VBM around the $\Gamma$ point.
Secondly, $D_{1}$ in $M(\vec{k})$ will lead to a gap at the $\Gamma$ point.
Thirdly, to reproduce band inversion, we need that $D_{1}m_{1}>0 \cap D_{1}n_{1}>0$.
More importantly, the dispersions of the model Hamiltonian given by Equation.~(2) are
$E(\vec{k})=f_{0}(\vec{k})\pm \sqrt{f_{1}^{2}(\vec{k})+f_{2}^{2}(\vec{k})+f_{3}^{2}(\vec{k})+f_{4}^{2}(\vec{k})+f_{5}^{2}(\vec{k})}$
and both dispersions are doubly degenerate.
As a result, a band crossing of this model requires $f_{1}=f_{2}=f_{3}=f_{4}=f_{5}=0$.
There are several discrete $\vec{k}$ points near the $\Gamma$
point may satisfy the above-mentioned conditions.
For example, $\vec{k}=(0,0,\pm \sqrt{-\frac{D_{2}}{E_{2}}})$
when $D_{2}E_{2}< 0 \cap \frac{D_{1}}{m_{1}}=-\frac{D_{2}}{E_{2}}$ stands.
It means that we may find some Dirac crossings at the first sight.
However, on one hand, $\frac{D_{1}}{m_{1}}=-\frac{D_{2}}{E_{2}}$
is a very rigorous condition and can not be obtained without other symmetries.
More importantly, on the other hand, we can explain that
$\vec{k}$ points lying in the plane $m_{{\Gamma}-M-L-A}$ must induce a gap in the following.
There are three generator operators existing in the nonsymmorphic space group P$\overline{3}$c1,
we sign the three-fold rotation axis around z axis, the glide plane located at x = 0,
and the inversion symmetry as R$_{3z}$, G$_{x}$ and P, respectively.
The operation of G$_{x}$ acts in both the real space $(x,y,z)$ and the spin space as
\begin{equation}\label{eq:eps}
\begin{split}
&G_{x}: (x,y,z) \rightarrow (-x,y,z+\frac{1}{2})\\
&G_{x}: (s_{x},s_{y},s_{z}) \rightarrow (s_{x},-s_{y},-s_{z}).\\
\end{split}
\end{equation}
Similar with the analysis in the work by Fang et al.~\cite{nlsm-fc-rapidcom},
we can easily find
\begin{equation}\label{eq:eps}
\begin{split}
&G_{x}*(P*T)=e^{-ik_{z}}(P*T)*G_{x}.\\
\end{split}
\end{equation}
On the mirror invariant plane $k_{x}=0$, the bands can be labeled by its $G_{x}$ eigenvalues.
When SOC is considered, we have
\begin{equation}\label{eq:eps}
\begin{split}
&G_{x}^{2}=-e^{-ik_{z}}\\
\end{split}
\end{equation}
the minus sigh is because $G_{x}^{2}$ includes a 2$\Pi$ rotation in the spin space,
which gives a -1 for a spin-$\frac{1}{2}$ system.
So the eigenvalue of $G_{x}$ is $\pm ie^{-i\frac{k_{z}}{2}}$.
The existence of both P and T will preserve all bands
locally degenerated at every $\vec{k}$ point in the BZ when SOC is considered,
and the degenerated bands are related to each other by P$\ast$T.
Suppose at $(0,k_{y},k_{z})$, a Bloch function $|\psi(\vec{k})\rangle$
is an eigenstate of $G_{x}$ with eigenvalue $ie^{-i\frac{k_{z}}{2}}$,
then its degenerate partner $P\ast T|\psi(\vec{k})\rangle$ under $G_{x}$,
\begin{equation}\label{eq:eps}
\begin{split}
G_{x}(P * T)|\psi(\vec{k})\rangle&=e^{-ik_{z}}(P * T)G_{x}|\psi(\vec{k})\rangle\\
                                 &=e^{-ik_{z}}(P * T)ie^{-i\frac{k_{z}}{2}}|\psi(\vec{k})\rangle\\
                                 &=-ie^{-i\frac{k_{z}}{2}}(P * T)|\psi(\vec{k})\rangle.
\end{split}
\end{equation}
It means that the degenerated bands on the $k_{x}=0$ plane have opposite $G_{x}$ eigenvalues,
and two sets of such doublet bands generally anticross,
i.e., the bands with the same $G_{x}$ eigenvalues hybridize and avoid crossing.
In other words, nodal lines near the $\Gamma$ point
in the $k_{x}=0$ plane (without SOC) will disappear in the whole BZ when SOC is considered,
and $G_{x}$ symmetry plays the key role of prohibiting the band crossing between CBM and VBM,
even though the gap is very small ($\approx$ 5 meV) as the $\mathbf{k}\cdot\mathbf{p}$ Hamiltonian in Equation(2) has showed us.
As a result, this node-line semimetal will become an insulator when SOC is considered.
With both TRS and IS in this system, we can easily calculate the $Z_{2}$ index by multiplying all the parities
of all the occupied bands at all time-reversal-invariant momenta (TRIMs) using the method by Fu and Kane~\cite{TIs-with-inversion-2007-FuLiang}.
The results are shown in Table.~\ref{table:parity-P-3c1},
which indicates the P$\bar{3}$c1 YH$_{3}$ is a strong TI with $Z_{2}=(1,000)$ when SOC is taken into consideration.
Nevertheless,
the effect of the SOC is negligible because the H atom is small and the Y atom is also not very heavy;
there is only one d electron in the Y atom.
We have performed calculations with SOC and found that the SOC induced gap along the nodal ring is very small (about 5 meV),
which indicates that the effect of SOC could be ignored and the characteristic of the nodal ring can be preserved.
\begin{table}
\centering
\caption{The parities product of all the occupied bands at the eight TRIMs for the P$\bar{3}$c1 phase of YH$_{3}$.}
   \begin{tabular}{l c c c c r}%
       \hline\hline
          \textbf{TRIM}         & $\Gamma$  &   3M       &  A    &    3L     ;  & total\\
           Parity               &    +      &    -       &  -    &     -     ;  &   -\\
       \hline\hline
  \end{tabular}
  \label{table:parity-P-3c1}
\end{table}

\begin{figure}[th]
\begin{center}
\includegraphics[width=0.75\textwidth]{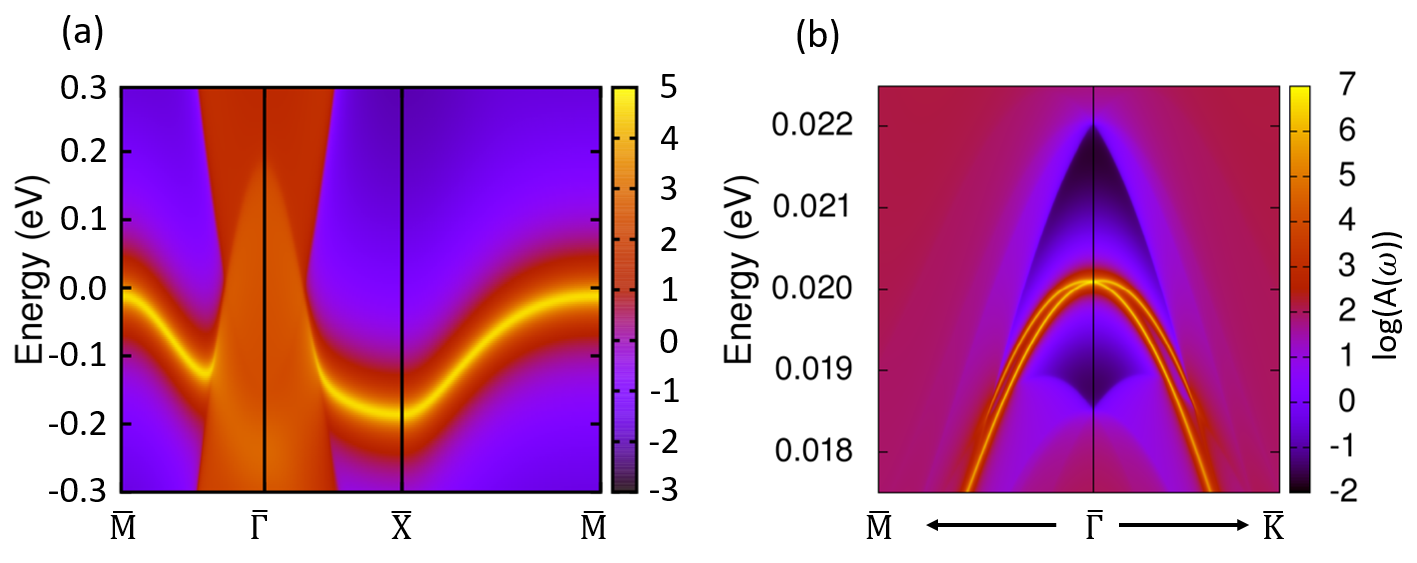}
\caption{%
(a)The surface states without SOC of P$\overline{3}$c1 YH$_{3}$ terminated with H atoms in the $\langle010\rangle$ direction,
(b)The surface states with SOC of P$\overline{3}$c1  YH$_{3}$ terminated with H atoms in the $\langle001\rangle$ direction.
}
\label{fig:ss}
\end{center}
\end{figure}

{\bf Surface states with and without SOC.}
Exotic topological surface states are an important property to identify various topological phases.
Based on the tight-binding model constructed with WLWFs and surface Green function methods~\cite{Qui-iter-sche,SurfaceGF,wu-wannier-tools-2017},
we have calculated the $\langle010\rangle$ surface states of the P$\bar{3}$c1 YH$_{3}$ without SOC
and the $\langle001\rangle$ surface states with SOC,
as shown in Fig.~\ref{fig:ss}(a) and Fig.~\ref{fig:ss}(b), respectively.
In particular, as the nice picture shows in Fig.~\ref{fig:ss}(a),
topological protected surface states without SOC signed with a bright curve
connects the nodal points across the boundary of BZ.
When SOC is included, the gap along $\Gamma \rightarrow A$ is so small ($\approx$ 5 meV)
that we may consider the CBM and VBM are nearly touched in the bulk band structure,
this phenomenon can be proved that we can find
the corresponding $\langle 001\rangle$ surface states connecting the psudo-touch points from Fig.~\ref{fig:ss}(b).
We think these topological protected surface states may be helpful
to identify the real ground state of YH$_{3}$ from those three candidates.
For example, we propose that
angle-resolved photoemission spectroscopy (ARPES) technique can be used to
investigate these surface states of this node-line semimetal candidate.
If a bright surface state could be found in the $\langle010\rangle$ direction,
and two parabolic bright curves with negative mass touching at the $\bar{\Gamma}$ point
could be found in the $\langle001\rangle$direction, 
then the YH$_{3}$ sample should be in the P$\bar{3}$c1 symmetry.

\section*{Conculsion}

In conclusion, based on first-principles calculations and effective model analysis,
we propose that the P$\overline{3}$c1 YH$_{3}$ is a nonsymmorphic symmetry protected node-line semimetal
when SOC is ignored.
This system has very clean electronic structures,
there are no other pockets except the ones composing the node line near the Fermi level.
The energy of the crossing points along the nodal loop has a very small energy range, from around -5 to 35 meV.
Therefore, this nodal loop is very ``flat'' in the energy and momentum space,
which makes YH$_{3}$ a perfect system for model analysis.
There are three node-lines related to each other by $R_{3z}$ symmetry
locating on three planes (signed as $m_{{\Gamma}-M-L-A}$) surrounding the $\Gamma$ point.
While SOC is taken into consideration,
the band crossings consisting of these node-lines will be gapped,
and the P$\overline{3}$c1 YH$_{3}$ becomes a strong topological insulator with Z$_{2}$ indices (1,000).
At last, we have calculated the surface states of this system to verify its topological properties.
We think our predictions should be helpful to identify the real ground state of YH$_{3}$ in experiments in the future.

\bibliographystyle{naturemag}

\begin{thebibliography}{10}
\expandafter\ifx\csname url\endcsname\relax
  \def\url#1{\texttt{#1}}\fi
\expandafter\ifx\csname urlprefix\endcsname\relax\def\urlprefix{URL }\fi
\providecommand{\bibinfo}[2]{#2}
\providecommand{\eprint}[2][]{\url{#2}}

\bibitem{TIs-Rmp-2010-Hasan}
\bibinfo{author}{Hasan, M.~Z.} \& \bibinfo{author}{Kane, C.~L.}
\newblock \bibinfo{title}{Colloquium : Topological insulators}.
\newblock \emph{\bibinfo{journal}{Rev. Mod. Phys.}}
  \textbf{\bibinfo{volume}{82}}, \bibinfo{pages}{3045} (\bibinfo{year}{2010}).

\bibitem{TIs-with-inversion-2011-(1057)FuLiang}
\bibinfo{author}{Qi, X.-L.} \& \bibinfo{author}{Zhang, S.-C.}
\newblock \bibinfo{title}{Topological insulators and superconductors}.
\newblock \emph{\bibinfo{journal}{Rev. Mod. Phys.}}
  \textbf{\bibinfo{volume}{83}}, \bibinfo{pages}{1057} (\bibinfo{year}{2011}).

\bibitem{DS-AB3-Wang-2012}
\bibinfo{author}{Wang, Z.} \emph{et~al.}
\newblock \bibinfo{title}{{D}irac semimetal and topological phase transitions
  in ${A}_{3}${B}i (${A}=${Na}, ${K}, ${Rb})}.
\newblock \emph{\bibinfo{journal}{Phys. Rev. B}} \textbf{\bibinfo{volume}{85}},
  \bibinfo{pages}{195320} (\bibinfo{year}{2012}).

\bibitem{3DDS-Cd3As2-Wang-2013}
\bibinfo{author}{Wang, Z.}, \bibinfo{author}{Weng, H.}, \bibinfo{author}{Wu,
  Q.}, \bibinfo{author}{Dai, X.} \& \bibinfo{author}{Fang, Z.}
\newblock \bibinfo{title}{Three-dimensional {D}irac semimetal and quantum
  transport in {C}d$_3${A}s$_2$}.
\newblock \emph{\bibinfo{journal}{Phys. Rev. B}} \textbf{\bibinfo{volume}{88}},
  \bibinfo{pages}{125427} (\bibinfo{year}{2013}).

\bibitem{3D-BiO2}
\bibinfo{author}{Young, S.~M.} \emph{et~al.}
\newblock \bibinfo{title}{{D}irac semimetal in three dimensions}.
\newblock \emph{\bibinfo{journal}{Phys. Rev. Lett.}}
  \textbf{\bibinfo{volume}{108}}, \bibinfo{pages}{140405}
  (\bibinfo{year}{2012}).

\bibitem{3dssm-design}
\bibinfo{author}{Gibson, Q.~D.} \emph{et~al.}
\newblock \bibinfo{title}{Three-dimensional {D}irac semimetals: Design
  principles and predictions of new materials}.
\newblock \emph{\bibinfo{journal}{Phys. Rev. B}} \textbf{\bibinfo{volume}{91}},
  \bibinfo{pages}{205128} (\bibinfo{year}{2015}).

\bibitem{Wan2011-weyl}
\bibinfo{author}{Wan, X.}, \bibinfo{author}{Turner, A.~M.},
  \bibinfo{author}{Vishwanath, A.} \& \bibinfo{author}{Savrasov, S.~Y.}
\newblock \bibinfo{title}{Topological semimetal and fermi-arc surface states in
  the electronic structure of pyrochlore iridates}.
\newblock \emph{\bibinfo{journal}{Phys. Rev. B}} \textbf{\bibinfo{volume}{83}},
  \bibinfo{pages}{205101} (\bibinfo{year}{2011}).

\bibitem{HgCrSe-2011}
\bibinfo{author}{Xu, G.}, \bibinfo{author}{Weng, H.}, \bibinfo{author}{Wang,
  Z.}, \bibinfo{author}{Dai, X.} \& \bibinfo{author}{Fang, Z.}
\newblock \bibinfo{title}{Chern semimetal and the quantized anomalous hall
  effect in {H}g{C}r${_2}${S}e$_4$}.
\newblock \emph{\bibinfo{journal}{Phys. Rev. Lett.}}
  \textbf{\bibinfo{volume}{107}}, \bibinfo{pages}{186806}
  (\bibinfo{year}{2011}).

\bibitem{TaAs-exper}
\bibinfo{author}{Lv, B.~Q.} \emph{et~al.}
\newblock \bibinfo{title}{Experimental discovery of weyl semimetal {TaAs}}.
\newblock \emph{\bibinfo{journal}{Phys. Rev. X}} \textbf{\bibinfo{volume}{5}},
  \bibinfo{pages}{031013} (\bibinfo{year}{2015}).

\bibitem{Xu2015-TaAs}
\bibinfo{author}{Xu, S.~Y.} \emph{et~al.}
\newblock \bibinfo{title}{{TOPOLOGICAL MATTER.} {D}iscovery of a {W}eyl fermion
  semimetal and topological {F}ermi arcs.}
\newblock \emph{\bibinfo{journal}{Science}} \textbf{\bibinfo{volume}{349}},
  \bibinfo{pages}{613} (\bibinfo{year}{2015}).

\bibitem{graphene-network}
\bibinfo{author}{Weng, H.} \emph{et~al.}
\newblock \bibinfo{title}{Topological node-line semimetal in three-dimensional
  graphene networks}.
\newblock \emph{\bibinfo{journal}{Phys. Rev. B}} \textbf{\bibinfo{volume}{92}},
  \bibinfo{pages}{045108} (\bibinfo{year}{2015}).

\bibitem{Cu3N-newZ2}
\bibinfo{author}{Kim, Y.}, \bibinfo{author}{Wieder, B.~J.},
  \bibinfo{author}{Kane, C.~L.} \& \bibinfo{author}{Rappe, A.~M.}
\newblock \bibinfo{title}{{D}irac {L}ine {N}odes in {Inversion-Symmetric}
  {C}rystals}.
\newblock \emph{\bibinfo{journal}{Phys. Rev. Lett.}}
  \textbf{\bibinfo{volume}{115}}, \bibinfo{pages}{036806}
  (\bibinfo{year}{2015}).

\bibitem{Cu3PdN}
\bibinfo{author}{Yu, R.}, \bibinfo{author}{Weng, H.}, \bibinfo{author}{Fang,
  Z.}, \bibinfo{author}{Dai, X.} \& \bibinfo{author}{Hu, X.}
\newblock \bibinfo{title}{Topological node-line semimetal and {D}irac semimetal
  state in antiperovskite {C}u$_3${P}d{N}}.
\newblock \emph{\bibinfo{journal}{Phys. Rev. Lett.}}
  \textbf{\bibinfo{volume}{115}}, \bibinfo{pages}{036807}
  (\bibinfo{year}{2015}).

\bibitem{Bian2016-PbTaSe2}
\bibinfo{author}{Bian, G.} \emph{et~al.}
\newblock \bibinfo{title}{Topological nodal-line fermions in spin-orbit metal
  {PbTaSe}$_2$}.
\newblock \emph{\bibinfo{journal}{Nat. Commun.}} \textbf{\bibinfo{volume}{7}},
  \bibinfo{pages}{10556} (\bibinfo{year}{2016}).

\bibitem{Bzdu2016Nodal}
\bibinfo{author}{Bzdusek, T.}, \bibinfo{author}{Wu, Q.},
  \bibinfo{author}{Ruegg, A.}, \bibinfo{author}{Sigrist, M.} \&
  \bibinfo{author}{Soluyanov, A.~A.}
\newblock \bibinfo{title}{Nodal-chain metals}.
\newblock \emph{\bibinfo{journal}{Nature}} \textbf{\bibinfo{volume}{538}},
  \bibinfo{pages}{75} (\bibinfo{year}{2016}).

\bibitem{nodalline-alkali-2016}
\bibinfo{author}{Li, R.} \emph{et~al.}
\newblock \bibinfo{title}{{D}irac {N}ode {L}ines in {P}ure {A}lkali {E}arth
  {M}etals}.
\newblock \emph{\bibinfo{journal}{Phys. Rev. Lett.}}
  \textbf{\bibinfo{volume}{117}}, \bibinfo{pages}{096401}
  (\bibinfo{year}{2016}).

\bibitem{Cd3As2-Liu-NM-2014}
\bibinfo{author}{Liu, Z.~K.} \emph{et~al.}
\newblock \bibinfo{title}{A stable three-dimensional topological {D}irac
  semimetal {C}d$_3${A}s$_2$.}
\newblock \emph{\bibinfo{journal}{Nat. Mater.}} \textbf{\bibinfo{volume}{13}},
  \bibinfo{pages}{677} (\bibinfo{year}{2014}).

\bibitem{Observation-Cd3As2-NC-XuSuYang-2014}
\bibinfo{author}{Neupane, M.} \emph{et~al.}
\newblock \bibinfo{title}{Observation of a three-dimensional topological
  {D}irac semimetal phase in high-mobility {C}d$_3${A}s$_2$}.
\newblock \emph{\bibinfo{journal}{{Nat. Commun.}}}
  \textbf{\bibinfo{volume}{{5}}}, \bibinfo{pages}{{3786}}
  (\bibinfo{year}{{2014}}).

\bibitem{Exp-realization-DS-Cava-2014}
\bibinfo{author}{Borisenko, S.} \emph{et~al.}
\newblock \bibinfo{title}{Experimental realization of a three-dimensional
  {D}irac semimetal}.
\newblock \emph{\bibinfo{journal}{Phys. Rev. Lett.}}
  \textbf{\bibinfo{volume}{113}}, \bibinfo{pages}{027603}
  (\bibinfo{year}{2014}).

\bibitem{Cd3As2:Evi-surface-state-Yi-2013}
\bibinfo{author}{Yi, H.} \emph{et~al.}
\newblock \bibinfo{title}{Evidence of topological surface state in
  three-dimensional {D}irac semimetal {Cd}$_3${As}$_2$.}
\newblock \emph{\bibinfo{journal}{Sci. Rep.}} \textbf{\bibinfo{volume}{4}},
  \bibinfo{pages}{6106} (\bibinfo{year}{2013}).

\bibitem{Landau-quan-CdAs-NM-Jeon-2014}
\bibinfo{author}{Jeon, S.} \emph{et~al.}
\newblock \bibinfo{title}{Landau quantization and quasiparticle interference in
  the three-dimensional {D}irac semimetal {Cd}$_3${As}$_2$.}
\newblock \emph{\bibinfo{journal}{Nat. Mater.}} \textbf{\bibinfo{volume}{13}},
  \bibinfo{pages}{851} (\bibinfo{year}{2014}).

\bibitem{Exp-verify-Na3Bi-Sci-2014}
\bibinfo{author}{Chen, Y.~L.} \emph{et~al.}
\newblock \bibinfo{title}{Discovery of a three-dimensional topological {D}irac
  semimetal, {Na}$_3${Bi}}.
\newblock \emph{\bibinfo{journal}{Science}} \textbf{\bibinfo{volume}{343}},
  \bibinfo{pages}{864} (\bibinfo{year}{2014}).

\bibitem{Exp-fermi-arc-Sci-XSY-2015}
\bibinfo{author}{Xu, S.~Y.} \emph{et~al.}
\newblock \bibinfo{title}{Observation of {F}ermi arc surface states in a
  topological metal.}
\newblock \emph{\bibinfo{journal}{Science}} \textbf{\bibinfo{volume}{347}},
  \bibinfo{pages}{294} (\bibinfo{year}{2015}).

\bibitem{TaAs-weyl-obser}
\bibinfo{author}{Huang, X.} \emph{et~al.}
\newblock \bibinfo{title}{Observation of the {Chiral-Anomaly-Induced}
  {N}egative {M}agnetoresistance in 3{D} {W}eyl {S}emimetal {TaAs}}.
\newblock \emph{\bibinfo{journal}{Phys. Rev. X}} \textbf{\bibinfo{volume}{5}},
  \bibinfo{pages}{031023} (\bibinfo{year}{2015}).

\bibitem{TaAs-weylnode-obs}
\bibinfo{author}{Lv, B.} \emph{et~al.}
\newblock \bibinfo{title}{Observation of {W}eyl nodes in {TaAs}}.
\newblock \emph{\bibinfo{journal}{Nat. Phys.}} \textbf{\bibinfo{volume}{11}},
  \bibinfo{pages}{724} (\bibinfo{year}{2015}).

\bibitem{spintexture-TaAs}
\bibinfo{author}{Lv, B.~Q.} \emph{et~al.}
\newblock \bibinfo{title}{Observation of {Fermi-Arc} {S}pin {T}exture in
  {TaAs}}.
\newblock \emph{\bibinfo{journal}{Phys. Rev. Lett.}}
  \textbf{\bibinfo{volume}{115}}, \bibinfo{pages}{217601}
  (\bibinfo{year}{2015}).

\bibitem{NbAs-weyldiscovery}
\bibinfo{author}{Xu, S.-Y.} \emph{et~al.}
\newblock \bibinfo{title}{Discovery of a {W}eyl fermion state with {F}ermi arcs
  in niobium arsenide}.
\newblock \emph{\bibinfo{journal}{Nat. Phys.}} \textbf{\bibinfo{volume}{11}},
  \bibinfo{pages}{748} (\bibinfo{year}{2015}).

\bibitem{TaP-2016observation}
\bibinfo{author}{Xu, N.} \emph{et~al.}
\newblock \bibinfo{title}{Observation of {W}eyl nodes and {F}ermi arcs in
  tantalum phosphide}.
\newblock \emph{\bibinfo{journal}{Nat. Commun.}} \textbf{\bibinfo{volume}{7}},
  \bibinfo{pages}{11006} (\bibinfo{year}{2016}).

\bibitem{WTe2-fermiarc-2016}
\bibinfo{author}{Bruno, F.~Y.} \emph{et~al.}
\newblock \bibinfo{title}{Observation of large topologically trivial {F}ermi
  arcs in the candidate type-{II} {W}eyl semimetal
  $\mathrm{WT}{\mathrm{e}}_{2}$}.
\newblock \emph{\bibinfo{journal}{Phys. Rev. B}} \textbf{\bibinfo{volume}{94}},
  \bibinfo{pages}{121112} (\bibinfo{year}{2016}).

\bibitem{WTe2-fermiarc2-2016}
\bibinfo{author}{Wu, Y.} \emph{et~al.}
\newblock \bibinfo{title}{Observation of {F}ermi arcs in the type-{II} {W}eyl
  semimetal candidate {WTe}$_2$}.
\newblock \emph{\bibinfo{journal}{Phys. Rev. B}} \textbf{\bibinfo{volume}{94}},
  \bibinfo{pages}{121113} (\bibinfo{year}{2016}).

\bibitem{MoTe2-2016experim}
\bibinfo{author}{Deng, K.} \emph{et~al.}
\newblock \bibinfo{title}{Experimental observation of topological {F}ermi arcs
  in type-{II} {W}eyl semimetal {MoTe}$_2$}.
\newblock \emph{\bibinfo{journal}{Nat. Phys.}} \textbf{\bibinfo{volume}{12}},
  \bibinfo{pages}{1105} (\bibinfo{year}{2016}).

\bibitem{nlsm-fc-rapidcom}
\bibinfo{author}{Fang, C.}, \bibinfo{author}{Chen, Y.}, \bibinfo{author}{Kee,
  H.-Y.} \& \bibinfo{author}{Fu, L.}
\newblock \bibinfo{title}{Topological nodal line semimetals with and without
  spin-orbital coupling}.
\newblock \emph{\bibinfo{journal}{Phys. Rev. B}} \textbf{\bibinfo{volume}{92}},
  \bibinfo{pages}{081201} (\bibinfo{year}{2015}).

\bibitem{ZrSiS-exp-2016}
\bibinfo{author}{Neupane, M.} \emph{et~al.}
\newblock \bibinfo{title}{Observation of topological nodal fermion semimetal
  phase in {ZrSiS}}.
\newblock \emph{\bibinfo{journal}{Phys. Rev. B}} \textbf{\bibinfo{volume}{93}},
  \bibinfo{pages}{201104} (\bibinfo{year}{2016}).

\bibitem{ZrSiSeTe-exp-2016}
\bibinfo{author}{Hu, J.} \emph{et~al.}
\newblock \bibinfo{title}{Evidence of {T}opological {Nodal-Line} {F}ermions in
  {ZrSiSe} and {ZrSiTe}}.
\newblock \emph{\bibinfo{journal}{Phys. Rev. Lett.}}
  \textbf{\bibinfo{volume}{117}}, \bibinfo{pages}{016602}
  (\bibinfo{year}{2016}).

\bibitem{class-TI-Tsc-2008}
\bibinfo{author}{Schnyder, A.~P.}, \bibinfo{author}{Ryu, S.},
  \bibinfo{author}{Furusaki, A.} \& \bibinfo{author}{Ludwig, A. W.~W.}
\newblock \bibinfo{title}{Classification of topological insulators and
  superconductors in three spatial dimensions}.
\newblock \emph{\bibinfo{journal}{Phys. Rev. B}} \textbf{\bibinfo{volume}{78}},
  \bibinfo{pages}{195125} (\bibinfo{year}{2008}).

\bibitem{TCI-FuLiang-2011}
\bibinfo{author}{Fu, L.}
\newblock \bibinfo{title}{Topological crystalline insulators}.
\newblock \emph{\bibinfo{journal}{Phys. Rev. Lett.}}
  \textbf{\bibinfo{volume}{106}}, \bibinfo{pages}{106802}
  (\bibinfo{year}{2011}).

\bibitem{SG-class-2013}
\bibinfo{author}{Slager, R.-J.}, \bibinfo{author}{Mesaros, A.},
  \bibinfo{author}{Juricic, V.} \& \bibinfo{author}{Zaanen, J.}
\newblock \bibinfo{title}{The space group classification of topological band
  insulators}.
\newblock \emph{\bibinfo{journal}{Nat. Phys.}} \textbf{\bibinfo{volume}{9}},
  \bibinfo{pages}{98} (\bibinfo{year}{2013}).

\bibitem{TCI-SG-2013}
\bibinfo{author}{Jadaun, P.}, \bibinfo{author}{Xiao, D.}, \bibinfo{author}{Niu,
  Q.} \& \bibinfo{author}{Banerjee, S.~K.}
\newblock \bibinfo{title}{Topological classification of crystalline insulators
  with space group symmetry}.
\newblock \emph{\bibinfo{journal}{Phys. Rev. B}} \textbf{\bibinfo{volume}{88}},
  \bibinfo{pages}{085110} (\bibinfo{year}{2013}).

\bibitem{TCI-TSC-2014}
\bibinfo{author}{Shiozaki, K.} \& \bibinfo{author}{Sato, M.}
\newblock \bibinfo{title}{Topology of crystalline insulators and
  superconductors}.
\newblock \emph{\bibinfo{journal}{Phys. Rev. B}} \textbf{\bibinfo{volume}{90}},
  \bibinfo{pages}{165114} (\bibinfo{year}{2014}).

\bibitem{classify-msym-2014-chiu}
\bibinfo{author}{Chiu, C.-K.} \& \bibinfo{author}{Schnyder, A.~P.}
\newblock \bibinfo{title}{Classification of reflection-symmetry-protected
  topological semimetals and nodal superconductors}.
\newblock \emph{\bibinfo{journal}{Phys. Rev. B}} \textbf{\bibinfo{volume}{90}},
  \bibinfo{pages}{205136} (\bibinfo{year}{2014}).

\bibitem{H-storage}
\bibinfo{author}{Sakintuna, B.}, \bibinfo{author}{Lamari-Darkrim, F.} \&
  \bibinfo{author}{Hirscher, M.}
\newblock \bibinfo{title}{Metal hydride materials for solid hydrogen storage: A
  review}.
\newblock \emph{\bibinfo{journal}{Int. J. Hydrogen Energy}}
  \textbf{\bibinfo{volume}{32}}, \bibinfo{pages}{1121} (\bibinfo{year}{2007}).

\bibitem{BaH-2007}
\bibinfo{author}{Smith, J.~S.}, \bibinfo{author}{Desgreniers, S.},
  \bibinfo{author}{Tse, J.~S.} \& \bibinfo{author}{Klug, D.~D.}
\newblock \bibinfo{title}{High-pressure phase transition observed in barium
  hydride}.
\newblock \emph{\bibinfo{journal}{J. Appl. Phys.}}
  \textbf{\bibinfo{volume}{102}}, \bibinfo{pages}{043520}
  (\bibinfo{year}{2007}).

\bibitem{SnH4-2007}
\bibinfo{author}{Tse, J.~S.}, \bibinfo{author}{Yao, Y.} \&
  \bibinfo{author}{Tanaka, K.}
\newblock \bibinfo{title}{Novel superconductivity in {M}etallic {SnH}$_4$ under
  {H}igh {P}ressure}.
\newblock \emph{\bibinfo{journal}{Phys. Rev. Lett.}}
  \textbf{\bibinfo{volume}{98}}, \bibinfo{pages}{117004}
  (\bibinfo{year}{2007}).

\bibitem{SiH3-2010}
\bibinfo{author}{Jin, X.} \emph{et~al.}
\newblock \bibinfo{title}{Superconducting high-pressure phases of disilane}.
\newblock \emph{\bibinfo{journal}{Proc. Natl. Acad. Sci. USA}}
  \textbf{\bibinfo{volume}{107}}, \bibinfo{pages}{9969} (\bibinfo{year}{2010}).

\bibitem{IH-2015}
\bibinfo{author}{Shamp, A.} \& \bibinfo{author}{Zurek, E.}
\newblock \bibinfo{title}{Superconducting high-pressure phases composed of
  hydrogen and iodine.}
\newblock \emph{\bibinfo{journal}{J. Phys. Chem. Lett.}}
  \textbf{\bibinfo{volume}{6}}, \bibinfo{pages}{4067} (\bibinfo{year}{2015}).

\bibitem{HTSC-cui2017}
\bibinfo{author}{Duan, D.} \emph{et~al.}
\newblock \bibinfo{title}{Structure and superconductivity of hydrides at high
  pressures}.
\newblock \emph{\bibinfo{journal}{Natl. Sci. Rev.}}
  \textbf{\bibinfo{volume}{4}}, \bibinfo{pages}{121} (\bibinfo{year}{2017}).

\bibitem{metalH}
\bibinfo{author}{Ashcroft, N.~W.}
\newblock \bibinfo{title}{Metallic hydrogen: A high-temperature
  superconductor?}
\newblock \emph{\bibinfo{journal}{Phys. Rev. Lett.}}
  \textbf{\bibinfo{volume}{21}}, \bibinfo{pages}{1748} (\bibinfo{year}{1968}).

\bibitem{metalH-alloy}
\bibinfo{author}{Ashcroft, N.~W.}
\newblock \bibinfo{title}{Hydrogen dominant metallic alloys: High temperature
  superconductors?}
\newblock \emph{\bibinfo{journal}{Phys. Rev. Lett.}}
  \textbf{\bibinfo{volume}{92}}, \bibinfo{pages}{187002}
  (\bibinfo{year}{2004}).

\bibitem{YH3-sc}
\bibinfo{author}{Kim, D.~Y.}, \bibinfo{author}{Scheicher, R.~H.} \&
  \bibinfo{author}{Ahuja, R.}
\newblock \bibinfo{title}{Predicted {High-Temperature} {S}uperconducting
  {S}tate in the {H}ydrogen-{D}ense {Transition-Metal} {H}ydride {YH}$_3$ at 40
  {K} and 17.7 {GPa}}.
\newblock \emph{\bibinfo{journal}{Phys. Rev. Lett.}}
  \textbf{\bibinfo{volume}{103}}, \bibinfo{pages}{077002}
  (\bibinfo{year}{2009}).

\bibitem{YH4-6-SC}
\bibinfo{author}{Li, Y.} \emph{et~al.}
\newblock \bibinfo{title}{Pressure-stabilized superconductive yttrium
  hydrides}.
\newblock \emph{\bibinfo{journal}{Sci. Rep.}} \textbf{\bibinfo{volume}{5}},
  \bibinfo{pages}{9948} (\bibinfo{year}{2015}).

\bibitem{YH10-liu}
\bibinfo{author}{Liu, H.}, \bibinfo{author}{Naumov, I.~I.},
  \bibinfo{author}{Hoffmann, R.}, \bibinfo{author}{Ashcroft, N.~W.} \&
  \bibinfo{author}{Hemley, R.~J.}
\newblock \bibinfo{title}{Potential high-tc superconducting lanthanum and
  yttrium hydrides at high pressure.}
\newblock \emph{\bibinfo{journal}{Proc. Natl. Acad. Sci. USA}}
  \textbf{\bibinfo{volume}{114}}, \bibinfo{pages}{6990} (\bibinfo{year}{2017}).

\bibitem{YH10-ma}
\bibinfo{author}{Peng, F.} \emph{et~al.}
\newblock \bibinfo{title}{Hydrogen clathrate structures in rare earth hydrides
  at high pressures: Possible route to room-temperature superconductivity}.
\newblock \emph{\bibinfo{journal}{Phys. Rev. Lett.}}
  \textbf{\bibinfo{volume}{119}}, \bibinfo{pages}{107001}
  (\bibinfo{year}{2017}).

\bibitem{BiX-monolayer}
\bibinfo{author}{Liu, C.-C.} \emph{et~al.}
\newblock \bibinfo{title}{Low-energy effective hamiltonian for giant-gap
  quantum spin hall insulators in honeycomb $x$-hydride/halide
  $(x=\mathrm{N}\char21{}\mathrm{Bi})$ monolayers}.
\newblock \emph{\bibinfo{journal}{Phys. Rev. B}} \textbf{\bibinfo{volume}{90}},
  \bibinfo{pages}{085431} (\bibinfo{year}{2014}).

\bibitem{soc-mat}
\bibinfo{author}{Feng, W.}, \bibinfo{author}{Liu, C.-C.}, \bibinfo{author}{Liu,
  G.-B.}, \bibinfo{author}{Zhou, J.-J.} \& \bibinfo{author}{Yao, Y.}
\newblock \bibinfo{title}{First-principles investigations on the berry phase
  effect in spin--orbit coupling materials}.
\newblock \emph{\bibinfo{journal}{Comput. Mater. Sci.}}
  \textbf{\bibinfo{volume}{112}}, \bibinfo{pages}{428} (\bibinfo{year}{2016}).

\bibitem{2d-NL}
\bibinfo{author}{Yang, B.}, \bibinfo{author}{Zhang, X.} \&
  \bibinfo{author}{Zhao, M.}
\newblock \bibinfo{title}{{D}irac node lines in two-dimensional lieb lattices}.
\newblock \emph{\bibinfo{journal}{Nanoscale}} \textbf{\bibinfo{volume}{9}},
  \bibinfo{pages}{8740} (\bibinfo{year}{2017}).

\bibitem{NLSM-sum}
\bibinfo{author}{Yang, S.~Y.} \emph{et~al.}
\newblock \bibinfo{title}{Symmetry demanded topological nodal-line materials}.
\newblock \emph{\bibinfo{journal}{arXiv:1707.04523v2}}  (\bibinfo{year}{2017}).

\bibitem{LAPW}
\bibinfo{author}{Sj\"ostedt, E.}, \bibinfo{author}{Nordstr\"om, L.} \&
  \bibinfo{author}{Singh, D.~J.}
\newblock \bibinfo{title}{An alternative way of linearizing the augmented
  plane-wave method}.
\newblock \emph{\bibinfo{journal}{Solid State Commun.}}
  \textbf{\bibinfo{volume}{114}}, \bibinfo{pages}{15} (\bibinfo{year}{2000}).

\bibitem{LAPW-prb}
\bibinfo{author}{Madsen, G. K.~H.}, \bibinfo{author}{Blaha, P.},
  \bibinfo{author}{Schwarz, K.}, \bibinfo{author}{Sj\"ostedt, E.} \&
  \bibinfo{author}{Nordstr\"om, L.}
\newblock \bibinfo{title}{Efficient linearization of the augmented plane-wave
  method}.
\newblock \emph{\bibinfo{journal}{Phys. Rev. B}} \textbf{\bibinfo{volume}{64}},
  \bibinfo{pages}{195134} (\bibinfo{year}{2001}).

\bibitem{wien2k}
\bibinfo{author}{Blaha, P.}, \bibinfo{author}{Schwarz, K.},
  \bibinfo{author}{Madsen, G.}, \bibinfo{author}{Kvasnicka, D.} \&
  \bibinfo{author}{Luitz, J.}
\newblock \emph{\bibinfo{title}{WIEN2k: An Augmented Plane Wave plus Local
  Orbitals Program for Calculating Crystal Properties}}
  (\bibinfo{publisher}{Karlheinz Schwarz}, \bibinfo{address}{Technische
  Universitaet Wien, Vienna}, \bibinfo{year}{2001}).

\bibitem{soc}
\bibinfo{author}{Kune\ifmmode~\check{s}\else \v{s}\fi{}, J.},
  \bibinfo{author}{Nov\'ak, P.}, \bibinfo{author}{Schmid, R.},
  \bibinfo{author}{Blaha, P.} \& \bibinfo{author}{Schwarz, K.}
\newblock \bibinfo{title}{Electronic structure of fcc th: Spin-orbit
  calculation with ${6p}_{1/2}$ local orbital extension}.
\newblock \emph{\bibinfo{journal}{Phys. Rev. B}} \textbf{\bibinfo{volume}{64}},
  \bibinfo{pages}{153102} (\bibinfo{year}{2001}).

\bibitem{mlwfs-1997}
\bibinfo{author}{Marzari, N.} \& \bibinfo{author}{Vanderbilt, D.}
\newblock \bibinfo{title}{Maximally localized generalized wannier functions for
  composite energy bands}.
\newblock \emph{\bibinfo{journal}{Phys. Rev. B}} \textbf{\bibinfo{volume}{56}},
  \bibinfo{pages}{12847} (\bibinfo{year}{1997}).

\bibitem{MLWFs-RMPhys}
\bibinfo{author}{Marzari, N.}, \bibinfo{author}{Mostofi, A.~A.},
  \bibinfo{author}{Yates, J.~R.}, \bibinfo{author}{Souza, I.} \&
  \bibinfo{author}{Vanderbilt, D.}
\newblock \bibinfo{title}{Maximally localized wannier functions: Theory and
  applications}.
\newblock \emph{\bibinfo{journal}{Rev. Mod. Phys.}}
  \textbf{\bibinfo{volume}{84}}, \bibinfo{pages}{1419} (\bibinfo{year}{2012}).

\bibitem{Mostofi2007wannier90}
\bibinfo{author}{Mostofi, A.~A.} \emph{et~al.}
\newblock \bibinfo{title}{wannier90 : A tool for obtaining maximally-localised
  wannier functions}.
\newblock \emph{\bibinfo{journal}{Comput. Phys. Commun.}}
  \textbf{\bibinfo{volume}{178}}, \bibinfo{pages}{685} (\bibinfo{year}{2008}).

\bibitem{Qui-iter-sche}
\bibinfo{author}{Sancho, M. P.~L.}, \bibinfo{author}{Sancho, J. M.~L.} \&
  \bibinfo{author}{Rubio, J.}
\newblock \bibinfo{title}{Quick iterative scheme for the calculation of
  transfer matrices: application to mo (100)}.
\newblock \emph{\bibinfo{journal}{J. Phys. F:Met. Phys.}}
  \textbf{\bibinfo{volume}{14}}, \bibinfo{pages}{1205} (\bibinfo{year}{1984}).

\bibitem{SurfaceGF}
\bibinfo{author}{Sancho, M. P.~L.}, \bibinfo{author}{Sancho, J. M.~L.},
  \bibinfo{author}{Sancho, J. M.~L.} \& \bibinfo{author}{Rubio, J.}
\newblock \bibinfo{title}{Highly convergent schemes for the calculation of bulk
  and surface green functions}.
\newblock \emph{\bibinfo{journal}{J. Phys. F:Met. Phys.}}
  \textbf{\bibinfo{volume}{15}}, \bibinfo{pages}{851} (\bibinfo{year}{1985}).

\bibitem{YD3-npd}
\bibinfo{author}{Udovic, T.~J.}, \bibinfo{author}{Huang, Q.} \&
  \bibinfo{author}{Rush, J.~J.}
\newblock \bibinfo{title}{Characterization of the structure of {YD}3 by neutron
  powder diffraction}.
\newblock \emph{\bibinfo{journal}{J. Phys. Chem. Solids}}
  \textbf{\bibinfo{volume}{57}}, \bibinfo{pages}{423} (\bibinfo{year}{1996}).

\bibitem{YD3-prb}
\bibinfo{author}{Udovic, T.~J.}, \bibinfo{author}{Huang, Q.},
  \bibinfo{author}{Erwin, R.~W.}, \bibinfo{author}{Hj\"orvarsson, B.} \&
  \bibinfo{author}{Ward, R. C.~C.}
\newblock \bibinfo{title}{Structural symmetry of {YD}$_3$ epitaxial thin
  films}.
\newblock \emph{\bibinfo{journal}{Phys. Rev. B}} \textbf{\bibinfo{volume}{61}},
  \bibinfo{pages}{12701} (\bibinfo{year}{2000}).

\bibitem{Pereierls-YH3-prl}
\bibinfo{author}{Wang, Y.} \& \bibinfo{author}{Chou, M.~Y.}
\newblock \bibinfo{title}{Peierls distortion in hexagonal {YH}$_3$}.
\newblock \emph{\bibinfo{journal}{Phys. Rev. Lett.}}
  \textbf{\bibinfo{volume}{71}}, \bibinfo{pages}{1226} (\bibinfo{year}{1993}).

\bibitem{LaH}
\bibinfo{author}{Pebler, A.} \& \bibinfo{author}{Wallace, W.~E.}
\newblock \bibinfo{title}{Crystal structures of some lanthanide hydrides}.
\newblock \emph{\bibinfo{journal}{J. Phys. Chem.}}
  \textbf{\bibinfo{volume}{66}}, \bibinfo{pages}{148} (\bibinfo{year}{1962}).

\bibitem{P-3c1-agree}
\bibinfo{author}{Remhof, A.} \emph{et~al.}
\newblock \bibinfo{title}{Hydrogen and deuterium in epitaxial {Y}(0001) films:
  {S}tructural properties and isotope exchange}.
\newblock \emph{\bibinfo{journal}{Phys. Rev. B}} \textbf{\bibinfo{volume}{59}},
  \bibinfo{pages}{6689} (\bibinfo{year}{1999}).

\bibitem{P-3c1-2006}
\bibinfo{author}{Fedotov, V.~K.}, \bibinfo{author}{Antonov, V.~E.},
  \bibinfo{author}{Bashkin, I.~O.}, \bibinfo{author}{Hansen, T.} \&
  \bibinfo{author}{Natkaniec, I.}
\newblock \bibinfo{title}{Displacive ordering in the hydrogen sublattice of
  yttrium trihydride}.
\newblock \emph{\bibinfo{journal}{J. Phys.: Condens. Matter}}
  \textbf{\bibinfo{volume}{18}}, \bibinfo{pages}{1593} (\bibinfo{year}{2006}).

\bibitem{YH3-P63-prl}
\bibinfo{author}{Kelly, P.~J.}, \bibinfo{author}{Dekker, J.~P.} \&
  \bibinfo{author}{Stumpf, R.}
\newblock \bibinfo{title}{Theoretical prediction of the structure of insulating
  {YH}$_3$}.
\newblock \emph{\bibinfo{journal}{Phys. Rev. Lett.}}
  \textbf{\bibinfo{volume}{78}}, \bibinfo{pages}{1315} (\bibinfo{year}{1997}).

\bibitem{YH3-P63-prl-neu1}
\bibinfo{author}{Udovic, T. J.}, \bibinfo{author}{Huang, Q. } \&
  \bibinfo{author}{Rush, J. J.}
\newblock \bibinfo{title}{Comment on ``Theoretical Prediction of the Structure of Insulating Y${\mathrm{H}}_{3}$''}.
\newblock \emph{\bibinfo{journal}{Phys. Rev. Lett.}}
  \textbf{\bibinfo{volume}{79}}, \bibinfo{pages}{2920} (\bibinfo{year}{1997}).

\bibitem{YH3-P63-prl-neu2}
\bibinfo{author}{Kelly, P. J.}, \bibinfo{author}{Dekker, J. P.} \&
  \bibinfo{author}{Stumpf, R.}
\newblock \bibinfo{title}{Kelly, Dekker, and Stumpf Reply:}.
\newblock \emph{\bibinfo{journal}{Phys. Rev. Lett.}}
  \textbf{\bibinfo{volume}{79}}, \bibinfo{pages}{2921} (\bibinfo{year}{1997}).

\bibitem{YH3-P-3c1-nat}
\bibinfo{author}{Udovic, T. J.}, \bibinfo{author}{Huang, Q. }, \bibinfo{author}{Santoro, A. } \&
  \bibinfo{author}{Rush, J. J.}
\newblock \bibinfo{title}{The nature of deuterium arrangements in YD3 and other rare-earth trideuterides}.
\newblock \emph{\bibinfo{journal}{Z. Kristallogr}}
  \textbf{\bibinfo{volume}{223}}, \bibinfo{pages}{697} (\bibinfo{year}{2008}).

\bibitem{TIs-with-inversion-2007-FuLiang}
\bibinfo{author}{Fu, L.} \& \bibinfo{author}{Kane, C.~L.}
\newblock \bibinfo{title}{Topological insulators with inversion symmetry}.
\newblock \emph{\bibinfo{journal}{Phys. Rev. B}} \textbf{\bibinfo{volume}{76}},
  \bibinfo{pages}{045302} (\bibinfo{year}{2007}).

\bibitem{wu-wannier-tools-2017}
\bibinfo{author}{Wu, Q.}, \bibinfo{author}{Zhang, S.}, \bibinfo{author}{Song,
  H.-F.}, \bibinfo{author}{Troyer, M.} \& \bibinfo{author}{Soluyanov, A.~A.}
\newblock \bibinfo{title}{{WannierTools}: An open-source software package for
  novel topological materials}.
\newblock \emph{\bibinfo{journal}{arXiv preprint arXiv:1703.07789v1}}
  (\bibinfo{year}{2017}).

\end{thebibliography}

\vspace{5mm}

\noindent{\bf Acknowledgements}\\
We thank the fruitful discussions with Jiawei Ruan, Huaiqiang Wang, Rui Wang, and Ying Xu.
This work is supported by the MOST of China (Grant Nos: 2016YFA0300404, 2015CB921202),
the National Natural Science Foundation of China (Grant Nos: 51372112 and 11574133),
NSF Jiangsu province (No. BK20150012), the Fundamental Research Funds for the Central Universities,
and Special Program for Applied Research on Super Computation of the NSFC-Guangdong Joint Fund (the second phase).
Part of the calculations were performed on the supercomputer in the HPCC of Nanjing University
and "Tianhe-2" at NSCC guangzhou.\\

\noindent{\bf Author contributions}\\
J.S. supervised the project.
D.S. and T.C. performed the calculations.
D.S. and J.S. analysed the results and wrote the manuscript.
All authors discussed the results, and commented on the manuscript.\\

\noindent{\bf Additional information}\\
Competing financial interests: The authors declare no competing financial interests.

\end{document}